\documentclass[a4paper]{jpconf}
\usepackage{graphicx}
\usepackage{latexsym}
\begin{document}
\title{Screening and conductance relaxations in insulating granular aluminium thin films}

\author{Julien Delahaye and Thierry Grenet}

\address{Institut N\'eel and Universit\'e Joseph Fourier, B.P. 166, F-38042 Grenoble C\'edex 9}

\ead{julien.delahaye@grenoble.cnrs.fr}

\begin{abstract}
We have recently found in insulating granular Al thin film a new experimental feature \cite{DelahayePRL11}, namely the existence of a conductance relaxation that is not sensitive to gate voltage changes. This conductance relaxation is related to the existence of a metallic-like screening in the film and can be used to estimate its characteristic length scale. In the present paper, we give some experimental details on how this feature was measured and present our first results on the screening length temperature dependence.
\end{abstract}

\section{Introduction}

Very slow conductance relaxations have been found during the last 20 years in some disordered insulators \cite{BenChorinPRB91,VakninPRB02,MartinezPRL97,GrenetEPJB03,GrenetEPJB07,Frydman}. They are characterized by two salient and related features: a slow decrease of the conductance after a quench of the samples at low temperature (4K) and a conductance "dip" in electrical field effect measurements.
The current and still partial understanding of these features is based on the electron glass hypothesis \cite{DaviesPRL82,GrunwaldJPC82,PollakSEM82}. According to theoretical studies \cite{TsigankovPRB03,MalikPRB04,KoltonPRB05,LebanonPRB05,SamozaPRL08,AmirPRL09}, a system of localized electrons with disorder and ill-screened interactions will need an infinite time to reach its equilibrium state at low temperature. The approach to equilibrium is associated with a decrease of the conductance, in qualitative agreement with the conductance relaxation observed after a quench at low temperature.

\begin{figure}[h]
\begin{minipage}{18pc}
\includegraphics[width=18pc]{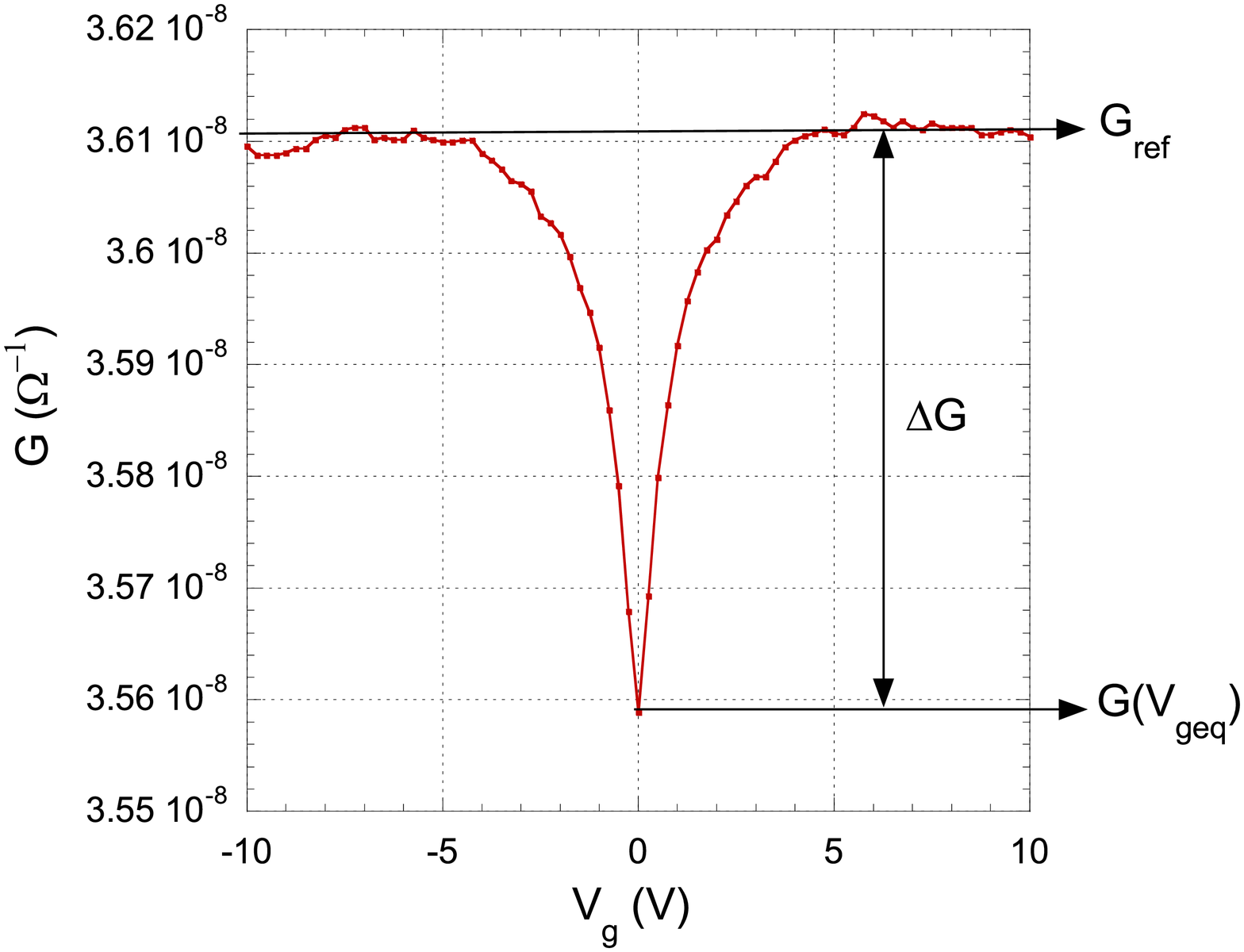}
\caption{\label{Figure1} $G(V_g)$ curve measured on a granular Al film after a quench at $4.2K$. The gate voltage $V_g$ was kept constant during 2 hours at $V_{geq} = 0V$ before sweeping $V_g$ between $-15V$ and $+15V$. The film was $20nm$ thick and $R_{\Box}=550M\Omega$ at $4.2K$.}
\end{minipage}\hspace{2pc}%
\begin{minipage}{18pc}
\includegraphics[width=18pc]{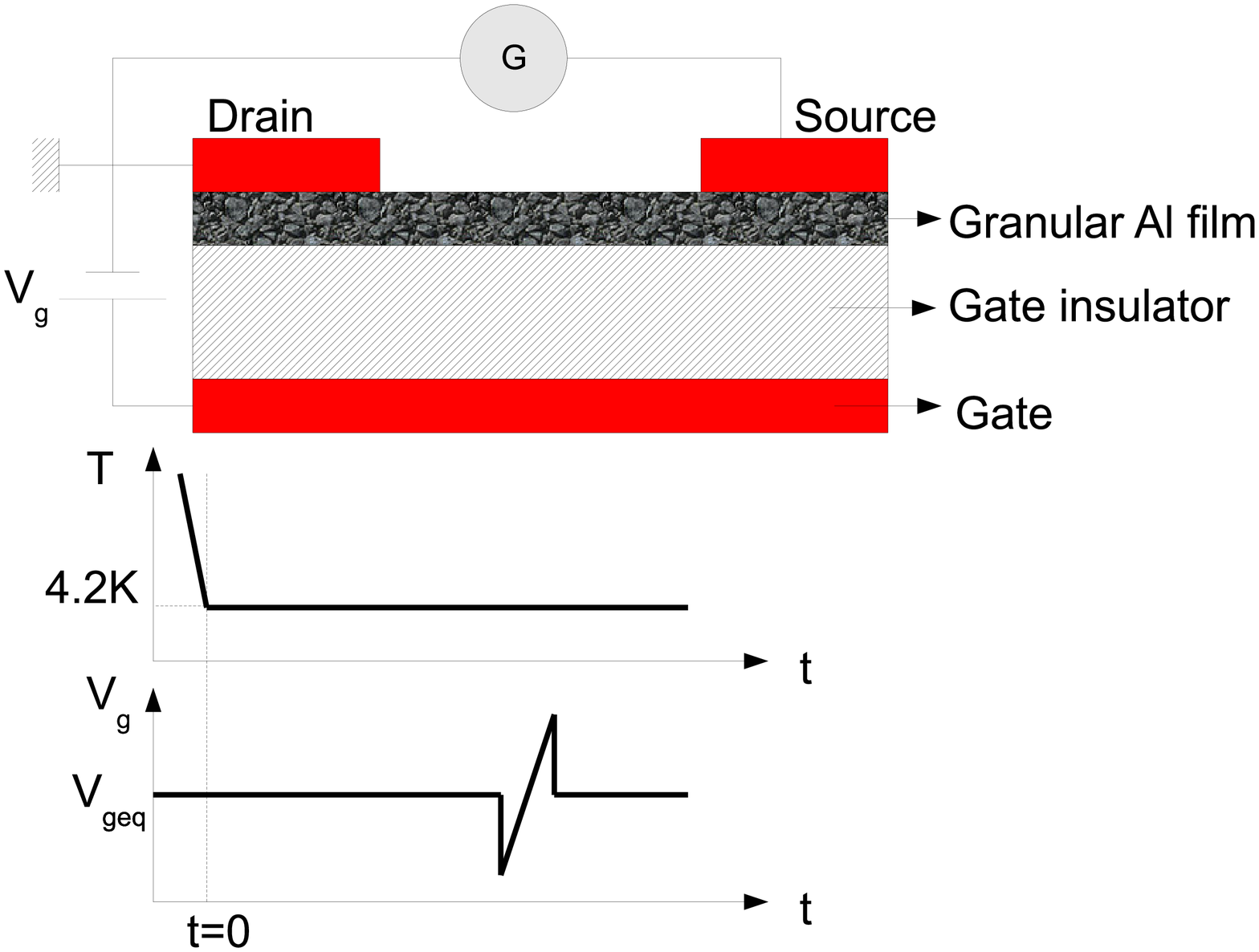}
\caption{\label{Figure2} Schematic drawing of a MOSFET device made of granular Al thin film and the protocol used to obtained the $G(V_g)$ curve of Fig. \ref{Figure1}.}
\end{minipage}
\end{figure}

Let's precise what the conductance "dip" is with the example of Fig. \ref{Figure1}. A MOSFET device which conducting channel is made of insulating granular Al thin film (see Fig. \ref{Figure2}) was cooled down to $4.2K$ and maintained under a fixed gate voltage $V_{geq} = 0V$ for 2 hours. The gate voltage was then swept around $V_{geq}$ at a constant scan rate and the $G(V_g)$ curve was recorded. A symmetrical dip, centered on $V_{geq}$, is clearly visible and reflects the partial equilibration of the system under $V_{geq}$. Out of the dip range, i.e for $|V_g|\geq 5V$, the conductance is roughly independent of the gate voltage and this baseline value $G_{ref}$ can be used to defined the conductance dip amplitude  $\Delta G = G_{ref} - G(V_{geq})$. $G_{ref}$ is also called (wrongfully, see below) the "off equilibrium" conductance since the system has never been allowed to equilibrate at these $V_g$'s. Contrary to our previous studies which were mainly focused on the time evolution of the conductance dip \cite{GrenetEPJB07}, we want to discuss here the time evolution of the baseline value $G_{ref}$. As we will see, it contains interesting information about the granular Al film properties.

\section{Experimentals}

The experimental challenge is to measure minute conductance variations of the granular Al channel at a fixed low T and over days. Since the relative amplitude of the conductance dip is of few percent or smaller, a relative precision of $\sim 10^{-2}\%$ is required.

The first problem is the T stability of the film during the measurement. Since the films are insulating, their resistance diverges exponentially when the T is lowered. $mK$ temperature variations in the liquid He range correspond typically to few \% changes in the conductance of the films, which is of the order of the variations we would like to measure as a function of time. The following set-up was thus employed. All the samples were mounted in a tight box and plunged into a $100l$ liquid He dewar. This box was either filled with some He exchange gas or let under vacuum. With He exchange gas, the quench to $4.2K$ was fast ($\simeq 10min$ from $90K$) but then the film follows the $mK$ variations of the liquid He bath around $4.2K$. A carbon glass thermometer close by was used to correct these T variations. Under vacuum, the quench to low T was much longer (few hours), but it was then possible to regulate the temperature with a high precision (better than $0.2mK$) by using the carbon glass thermometer and a resistive heater. The available temperatures in that case are in the range $5-15K$.

The second problem is the sensitivity of the electronic devices (lock-in, current amplifier, etc.) to room temperature drifts. For example, the current amplifier gain in the range $10^6$ - $10^8$ has a T drift of $100ppm/^o C$. We specially built a thermalized chamber with a T stability better than $0.1K$ in which all the electronic devices were placed. Doing so, we were able to detect conductance variations as small as $10^{-3}\%$ over weeks of measurements.

We have exclusively used granular Al films made by electron gun evaporation of Al under a partial pressure of $O_2$, as described elsewhere \cite{GrenetEPJB07}. In order to build MOSFET devices, granular Al films were deposited on top of heavily doped $Si$ wafers (the gate) covered by a $100nm$ thick thermally grown $SiO_2$ layer. The film conductance was measured by using a two terminal AC technique, employing a FEMTO current amplifier DLPCA 200, and a lock-in amplifier SR 7265. The source-drain voltage was such that G stays in the ohmic regime.

\section{The baseline relaxation}

In order to measure the baseline $G_{ref}$ time variations, we follow a simple experimental protocol. The MOSFET devices are cooled down from room T to liquid He (or to a T slightly above) and once the T has reached the desired value, we measure $G(V_g)$ curves around $V_{geq}$ at constant time interval. The time between two sweeps is kept much larger than the sweep time, usually $6000s$ between two $G(V_g)$ curves and $250s$ for a $-15V / +15V$ sweep.

A typical result for a $20nm$ thick film quenched with He exchange gas is shown in the Fig. \ref{Figure3} below. Without any T corrections, the observed baseline conductance variations of few $\%$ mainly reflect the mK fluctuations of the He bath. Interestingly, the dip amplitude is less sensitive to T so that T variations are small enough to allow a measurement of its time development. This is illustrated in the Fig. \ref{Figure4} where $\Delta G$ is plotted as a function of the time elapsed after the quench. The $\ln(t)$ dependence reported and discussed in previous studies \cite{GrenetEPJB07} is clearly visible.

\begin{figure}[h]
\begin{minipage}{18pc}
\includegraphics[width=18pc]{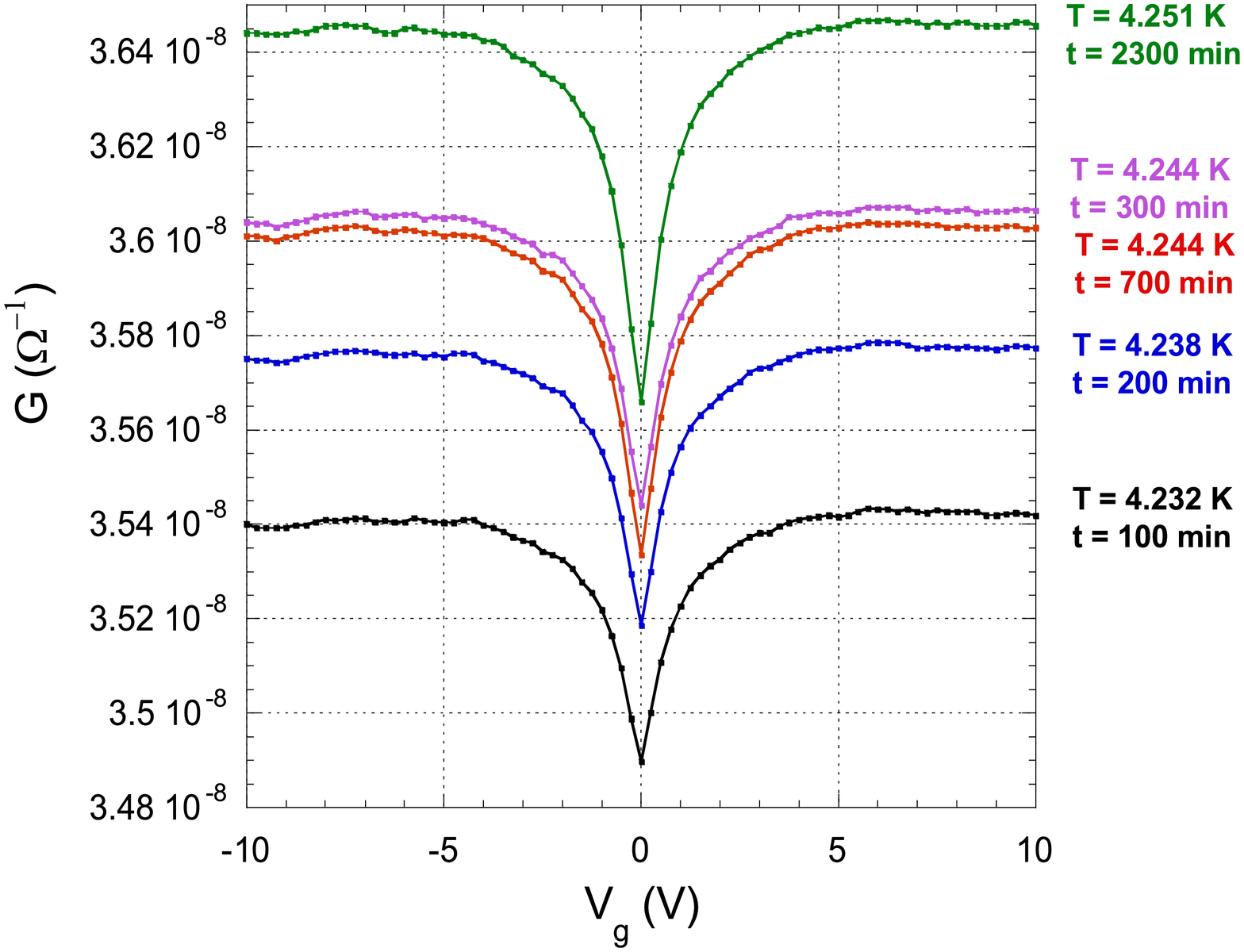}
\caption{\label{Figure3} $G(V_g)$ curves measured at different times t after a cool down to $4.2K$, without any corrections of the He bath T fluctuations. $V_{geq}=0V$ between $V_g$ sweeps. Same film as in Fig. \ref{Figure1}.}
\end{minipage}\hspace{2pc}%
\begin{minipage}{18pc}
\includegraphics[width=18pc]{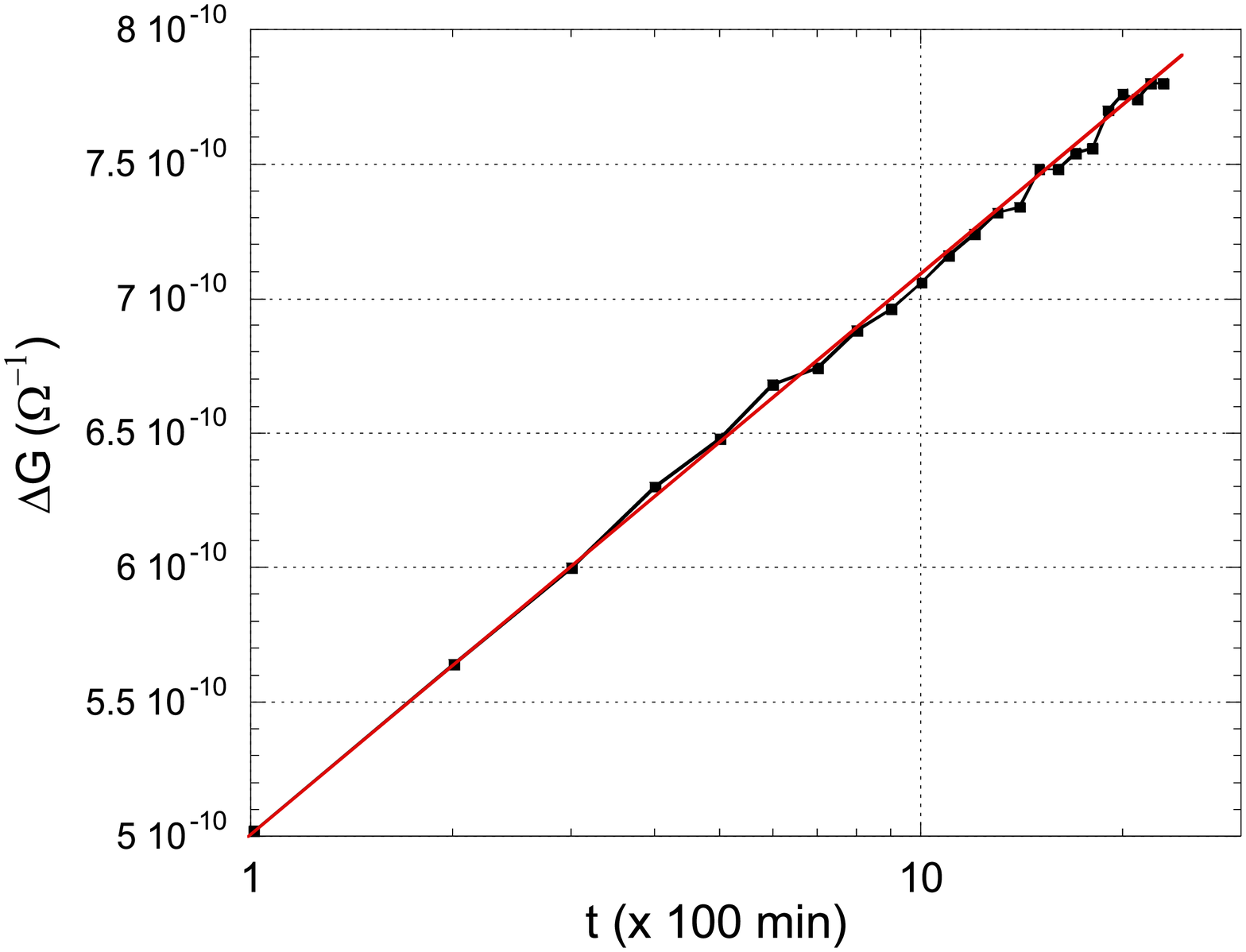}
\caption{\label{Figure4} Corresponding amplitude of the conductance dip $\Delta G(t)=G_{ref}-G(V_{geq})$.}
\end{minipage}
\end{figure}

\begin{figure}[h]
\begin{minipage}{18pc}
\includegraphics[width=18pc]{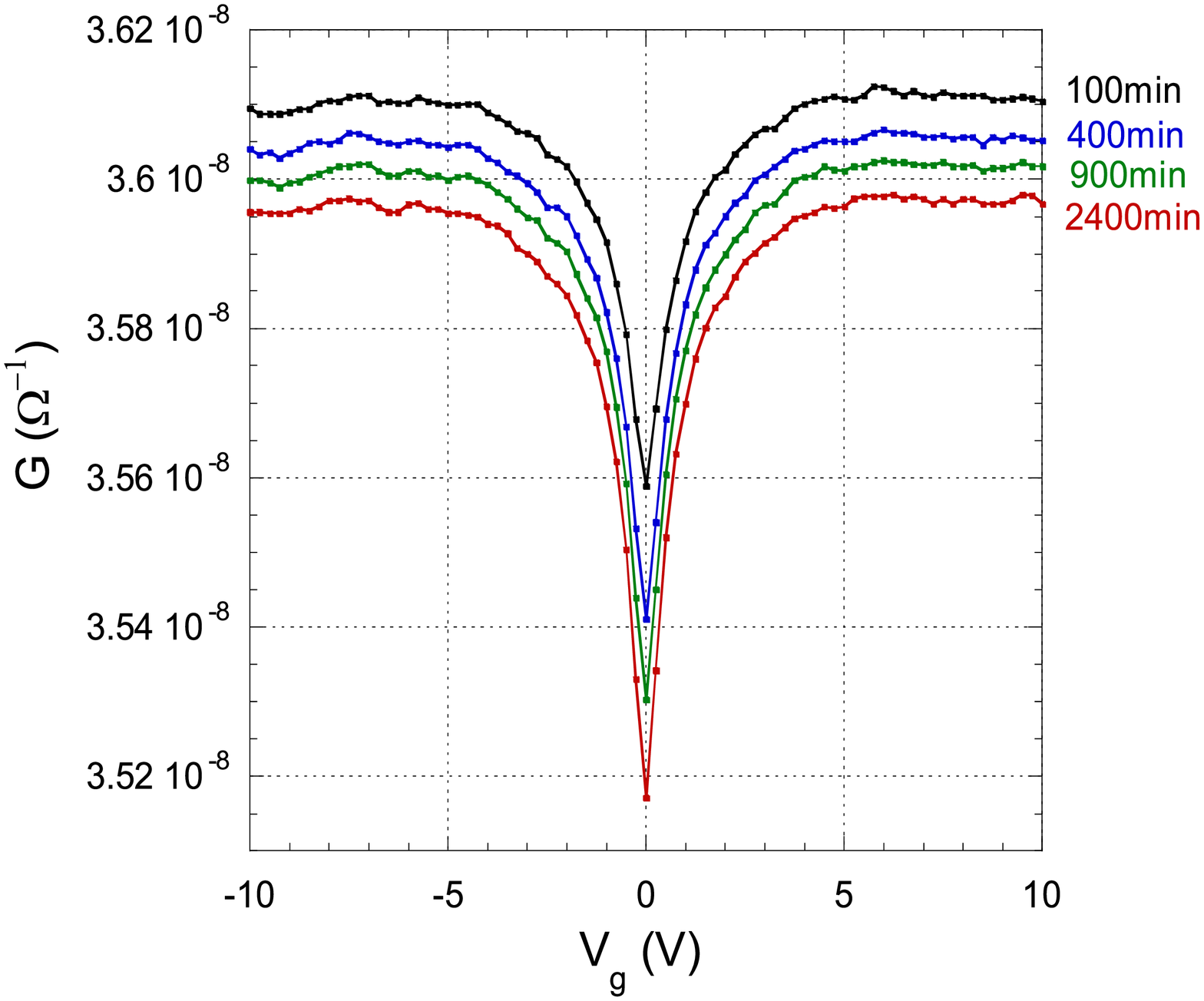}
\caption{\label{Figure5} $G(V_g)$ curves measured at different times after a cool down to $4.2K$ with a correction of the T variations. Same film as in Fig. \ref{Figure1}.}
\end{minipage}\hspace{2pc}%
\begin{minipage}{18pc}
\includegraphics[width=18pc]{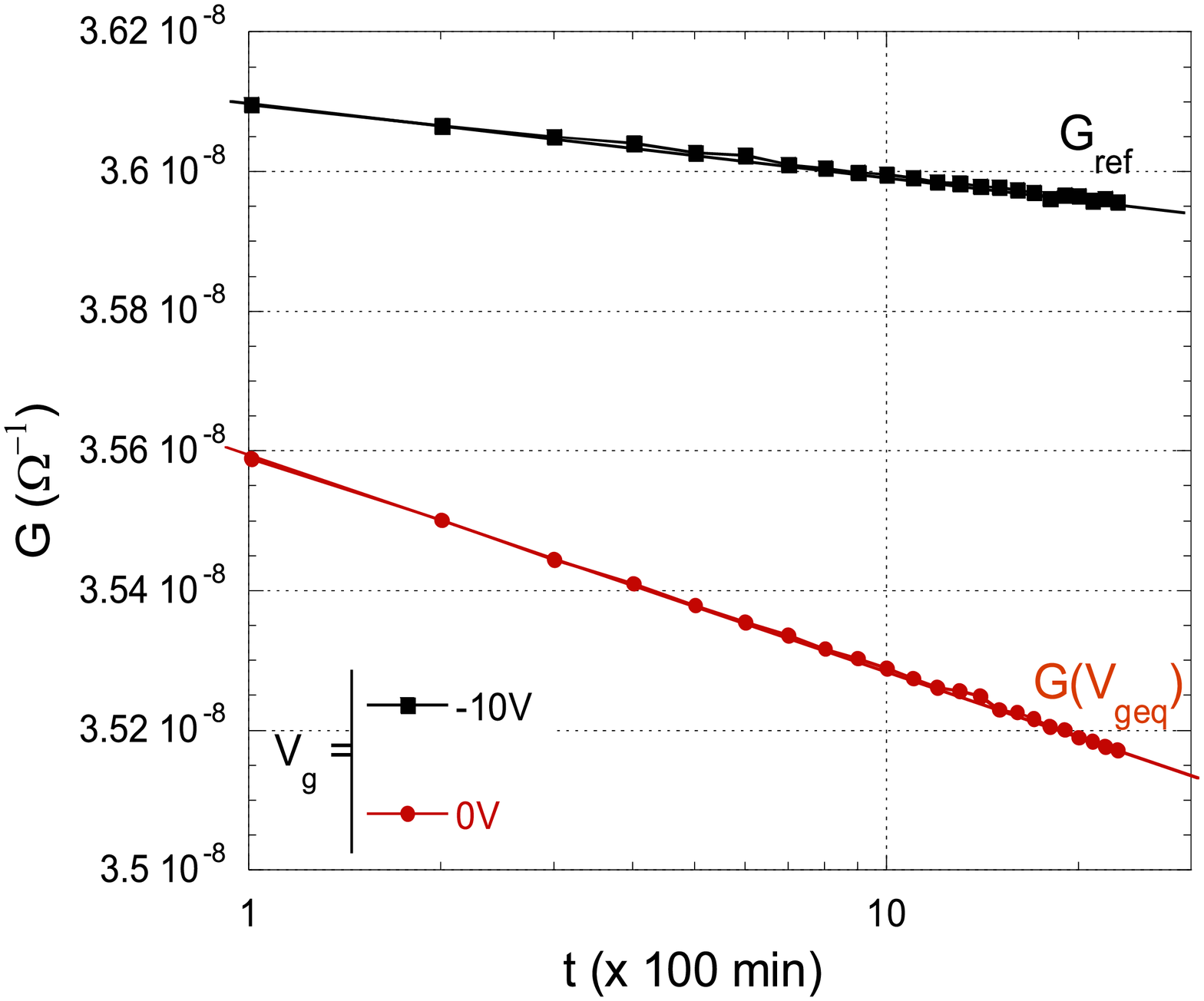}
\caption{\label{Figure6} Corresponding $G(t)$ curves at $V_{geq}$ ($V_{geq}=0V$) and out of the conductance dip region ($V_g=-10V$).}
\end{minipage}
\end{figure}

If we now correct for the T variations of the He bath (we can measure after a long time the G-T dependence), we get the curves shown in Fig. \ref{Figure5} and \ref{Figure6}. Both the baseline and the conductance minimum at $V_{geq}$ decrease with a $\ln(t)$ dependence. It will be noted that the curve of the Fig. \ref{Figure4} is simply the difference of the two $\ln(t)$ curves of the Fig. \ref{Figure6}.

\begin{figure}[h]
\begin{minipage}{18pc}
\includegraphics[width=18pc]{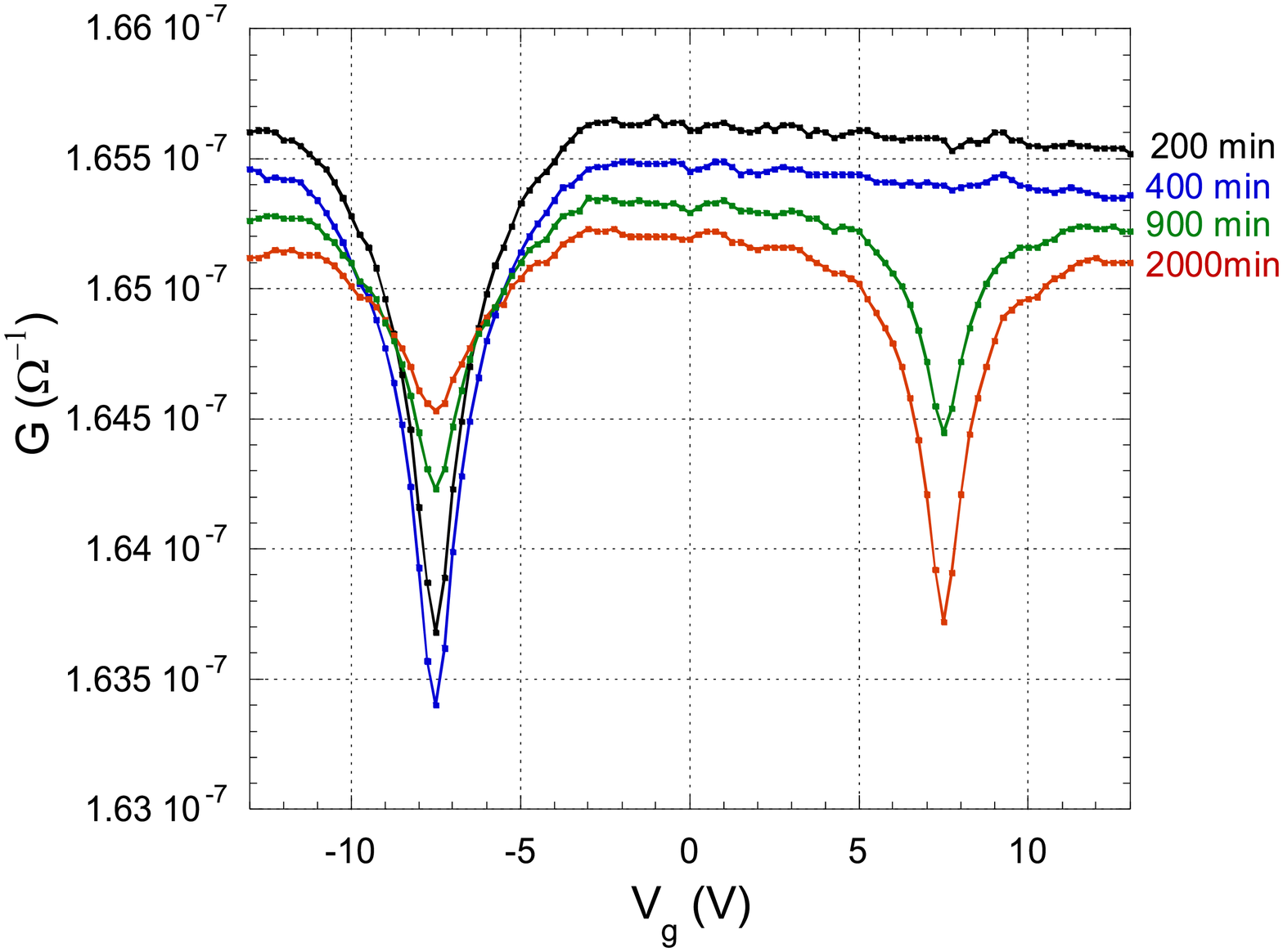}
\caption{\label{Figure7} $G(V_g)$ curves measured at different times after the cool down at $5.3K$. $V_{geq}$ was first set to $V_{geq1}=-7.5V$ and changed to $V_{geq2}=+7.5V$ after 800 min. Same film as in Fig. \ref{Figure1}.}
\end{minipage}\hspace{2pc}%
\begin{minipage}{18pc}
\includegraphics[width=18pc]{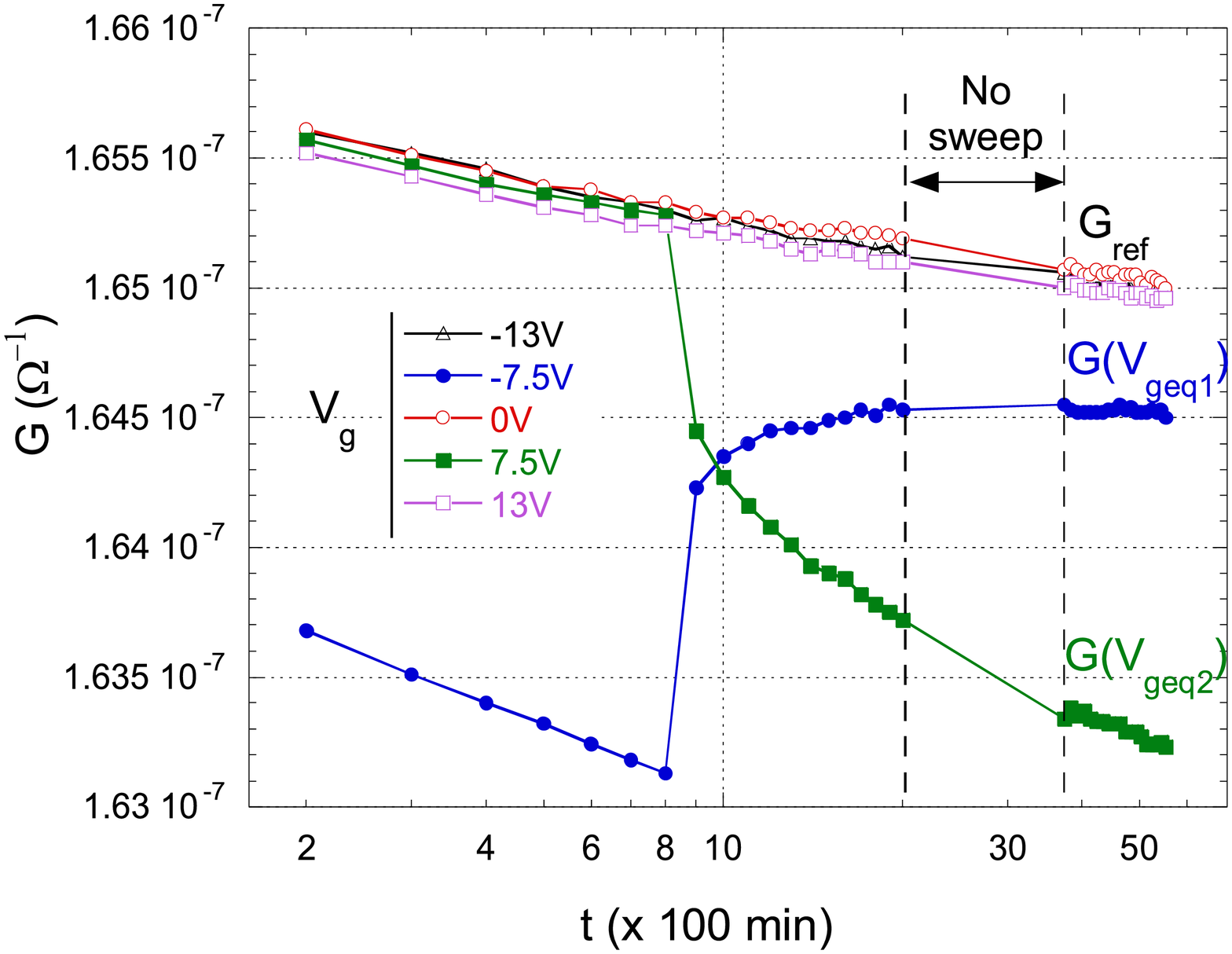}
\caption{\label{Figure8} Corresponding $G(t)$ curves at different $V_g$ values. The $V_g$ sweeps were suspended during 29 hours, 33 hours after the cool down.}
\end{minipage}
\end{figure}

We checked that the $V_{geq}$ values has no effect on the baseline decrease. Quantitatively similar $G_{ref}$ relaxations are observed in Fig. \ref{Figure7} and \ref{Figure8} when $V_{geq}$ was first set to $V_{geq1}=-7.5V$ and then change to $V_{geq2}=+7.5V$ 800 min after a quench to $5.3K$. Two dips are visible in the $G(V_g)$ curves, but for $V_g$ values far enough from $V_{geq1}$ and $V_{geq2}$, a regular $\ln(t)$ decrease of the conductance is obtained. During this run of measurement, we have also suspended the $V_g$ sweeps during 30 hours. No significant change in the baseline decrease is found which exonerates the $V_g$ sweeps in the G relaxations.

\section{Discussion}

The $\ln(t)$ dependence of $G(V_{geq})$ reflects the well known approach of the system to equilibrium after a quench \cite{VakninPRB02,GrenetEPJB07}. But how can we understand the $\ln(t)$ relaxation of the baseline?  In Ref. \cite{DelahayePRL11}, we described in details how a metallic-like screening length $L_{sc}$ smaller than the thickness $T_h$ of the films ($20nm$ for the previous figures) can provide a simple explanation to the observed baseline relaxation. For a given $V_g$, only the part of the film which is at a distance smaller than $L_{sc}$ from the gate insulator will feel an electric field (let's call it the "unscreened" layer). The field in the other part of the film (the "screened" layer) will be much smaller. If the screening is much faster than the typical time scale of the $V_g$ changes ($\simeq 1s$), there won't be any electrostatic perturbation during a $V_g$ sweep in the "screened" layer. In other words, only the "unscreened" layer is pushed out of "equilibrium" by the $V_g$ sweeps. By assuming a parallel conduction for the different layers, we can say that the $\ln (t)$ dependence of $G(V_{geq})$ reflects the relaxation of the whole film ("screened" and "unscreened" layers) towards equilibrium, whereas the $\ln(t)$ dependence of $G_{ref}$ reflects the relaxation of the "screened" layer alone (the relaxation of the "unscreened" layer is given by the conductance dip growth). It was shown in Ref. \cite{DelahayePRL11} that the slope ratio $SR$ of the $G(V_{geq}) - \ln(t)$ slope to the $G_{ref} - \ln(t)$ slope is given by:
\begin{equation}\label{SR}
    SR = T_h/(T_h - L_{sc})
\end{equation}
 and can thus be used to estimate the $L_{sc}$ value. In the Fig. \ref{Figure6}, $SR$ is equal to 2.3, which gives $L_{sc}=13nm$ ($T=4.2K$). We have measured this slope ratio at $4.2K$ on other 20nm thick granular Al thin films with square resistances $R_{\Box}$ between $1M\Omega$ and $10G\Omega$. The slope ratios were all found between 2 and 3, corresponding to quite similar $L_{sc}$ values in the range $10-13nm$. This $L_{sc}$ estimate of $\simeq10nm$ at $4K$ was confirmed by the study of the $G(V_g)$ relaxations for granular Al films of different thicknesses \cite{DelahayePRL11}. It lies between the percolation radius $L_0$, estimated to be about $20-40nm$ at $4.2K$ (see below), and the typical grain size of Al, which according to TEM studies is in the range $2-4nm$.

The existence of a metallic-like screening length in our insulating films is not surprising. The films have a low but finite conductivity at low temperature. Some electrons can thus move by hopping through the whole sample and screen an electric field. Other electrons that remain localized during the time scale of the experiment give a dielectric-like screening contribution (polarization). A transition between a metallic screening at high T, where most of the electrons are diffusive, and a slow dielectric response at low T, where most of the electrons remain localized in finite size clusters was indeed found in numerical studies of disordered insulators \cite{XuePRB88,DiazPRB99,KoltonPRB05}.

\begin{figure}[h]
\begin{minipage}{18pc}
\includegraphics[width=18pc]{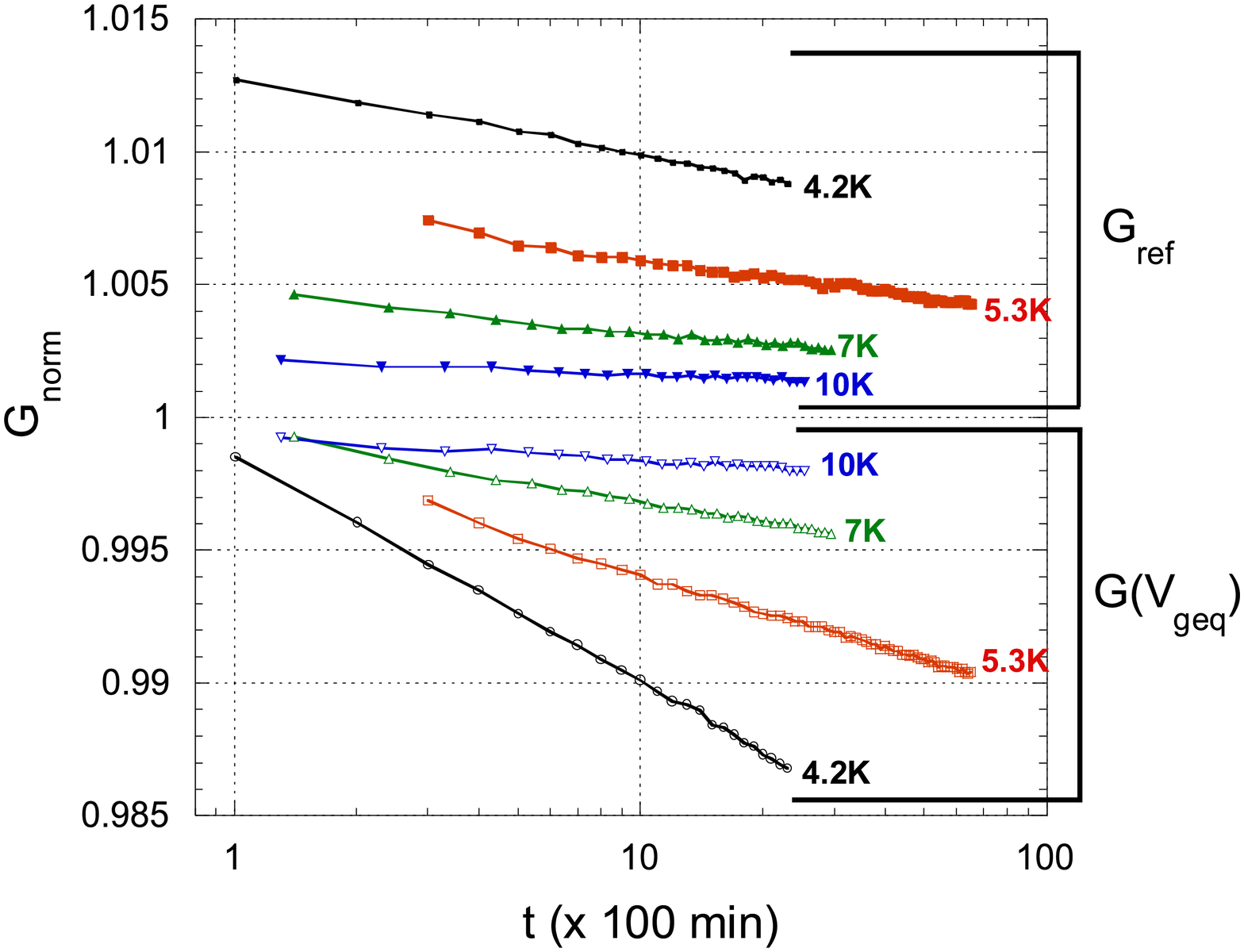}
\caption{\label{Figure9} Normalized conductance relaxations measured at $-10V$ ($G_{ref}$) and at $V_{geq} = 0V$ ($G(V_{geq}$)) as a function of time and after the cool down at different temperatures. Same film as in Fig. \ref{Figure1}.}
\end{minipage}\hspace{2pc}%
\begin{minipage}{18pc}
\includegraphics[width=18pc]{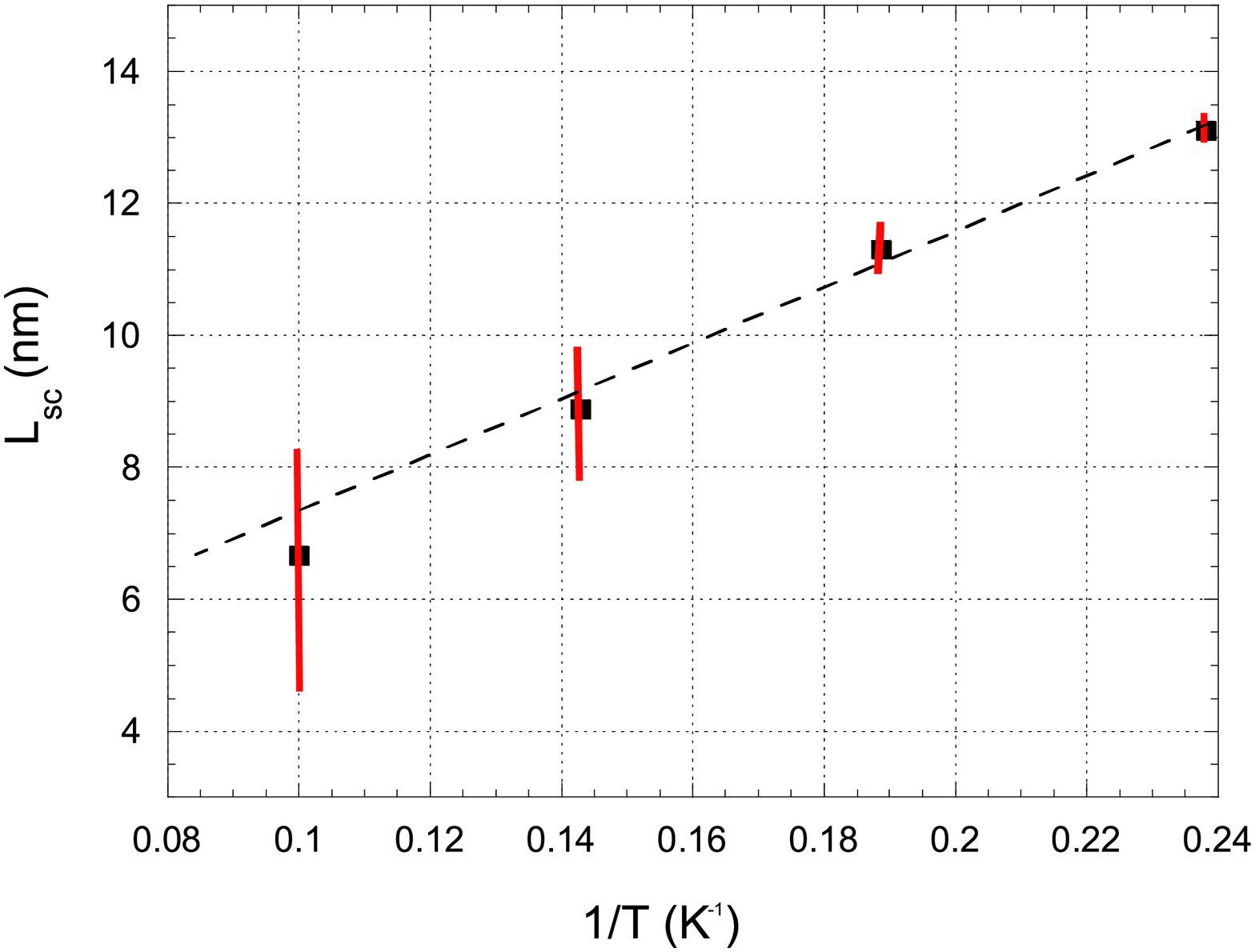}
\caption{\label{Figure10} The screening length values extracted from the slope ratios SR of Fig. \ref{Figure9} plotted versus $1/T$.}
\end{minipage}
\end{figure}

One may expect that the frequency and the temperature dependence of the screening properties are different in our hopping system from the metal case. By measuring the slope ratio SR at different temperatures between $4.2K$ and $10K$ for the 20nm thick film of the previous figures (see Fig. \ref{Figure9} and \ref{Figure10}), we found that $L_{sc}$ decreases as the T (and thus G) increases. This increase can be described by a $1/T$ dependence predicted by some theoretical studies \cite{MullerPRB07}, but the uncertainty is large.

We do not see any evidence for a large change with time in the effective screening length over the time scales probed by our experiment (from $1s$ to few days). Since glassy effects represent only conductance changes of few percent or less, their contribution to the screening might also be small and not detectable by our experiments. According to theoretical studies of the electron glass, the single electrons hops are believed to be fast (Maxwell time) and could be responsible for the short time screening length measured here. Much slower multi-electron hops will result in a slow decrease of the screening length as a function of time \cite{MullerPRB07}.

In light of these findings, some of the results that were published before have to be quantitatively reconsidered. It is indeed the case of our percolation radius ($L_0$) estimates obtained from the conductance fluctuations of $20nm$ thick mesoscopic films \cite{DelahayeEPJB08}. Since only the film layer located at a distance smaller than $L_{sc}$ to the gate insulator is sensitive to $V_g$ changes, the $L_0$ values published in Ref. \cite{DelahayeEPJB08} must thus be divided by $\sqrt{(T_h/L_{sc})}=\sqrt 2$. Due to the T dependence of the screening length highlighted in Fig. \ref{Figure10}, the actual decrease of the relative conductance dip amplitude $\Delta G/G$ as T is increased is weaker than it was published in Ref. \cite{GrenetEPJB07}.

Finally, we should say that no baseline relaxation was found in indium oxide films where similar measurements to ours have been made. Published data concern films of 20nm thick or less, and this will suggest a screening length larger than 20nm. It would thus be interesting to measure thicker films to see if the screening length limit can be attained. Such experiments are obviously not possible for ultrathin films of metals, in which the insulating character is due to their thinness.

\section{Conclusion}
In conclusion, we have demonstrated in our granular Al insulating films the existence of a conductance relaxation which is undisturbed by a gate voltage change. This relaxation is observed after a quench at low temperature and obeys a $\ln(t)$ dependence. According to Ref. \cite{DelahayePRL11}, the comparison with the $\ln(t)$ increase of the conductance dip amplitude allows an estimate of the metallic-like screening length of the films. This length is found to be of the order of $10nm$ at $4K$ and to decrease as the temperature is increased.

\ack

This research has been supported by the French-National Research Agency ANR (Contract 05-JC05-44044), the R\'egion Rh\^ones Alpes (CIBLE 2010 program) and the Universit\'e Joseph Fourier (SMINGUE 2010 program). Discussions with A. Frydman and M. Pollak are gratefully acknowledged.

\end{document}